\documentclass[%
 reprint,
 amsmath,amssymb,
 aps,
 showkeys,
]{revtex4-2}

\usepackage{graphicx}
\usepackage{dcolumn}
\usepackage{bm}
\usepackage[caption=false]{subfig}

\begin{document}

\preprint{APS/123-QED}

\title{Lattice patch structure for fixed-frequency transmon quantum computer with high-fidelity CNOT gates}

\author{Chanpyo Kim}
\email{freezeticket@hanyang.ac.kr}
\affiliation{Department of Applied Physics, Hanyang University, Ansan, South Korea}

\author{Jeongsoo Kang}
\email{jskang1202@hanyang.ac.kr}
\affiliation{Department of Applied Physics, Hanyang University, Ansan, South Korea}

\author{Younghun Kwon}
\email{yyhkwon@hanyang.ac.kr}
\thanks{Corresponding author} 
\affiliation{Department of Applied Physics, Hanyang University, Ansan, South Korea}

\date{\today}

\begin{abstract}
Superconducting transmon processors represent a leading platform for large-scale quantum computing due to their high gate fidelities and scalability. However, conventional qubit-coupler-qubit (QCQ) architectures face critical physical and structural bottlenecks, notably frequency crowding [spectator qubit collisions] during system scaling and inefficient mapping onto the standard surface code.To overcome these limitations, we propose a novel lattice-patch architecture that couples four fixed-frequency transmons to a single fixed-frequency coupler.This design enhances qubit connectivity and maps directly onto the surface-code lattice unit [plaquette], thereby minimizing the compilation overhead associated with logical qubit implementation. Furthermore, utilizing an entirely fixed-frequency design intrinsically eliminates susceptibility to external flux noise, ensuring robust operational stability.Multi-level numerical simulations demonstrate CNOT gate fidelities exceeding 0.98 across all six connectivity directions within the patch. Nevertheless, the complex interaction network of the four-qubit architecture induces unintended residual phase accumulation during cross-resonance driving. This parasitic effect necessitates precise calibration, achievable via virtual $R_z$ gates [software phase updates]. Ultimately, our results establish the lattice-patch architecture as an efficient, robust building block for future fault-tolerant quantum computers.
\end{abstract}


\maketitle

\section{Introduction\label{sec:Introduction}}
test test

A variety of quantum computing algorithms, including Shor's algorithm and Grover's algorithm, have demonstrated both the necessity and the feasibility of implementing quantum computers. In particular, the potential of quantum computing to surpass the limits of classical computation has driven its rapid adoption across diverse industrial and research fields \cite{shor_polynomial-time_nodate, grover_fast_1996, bae_generalized_2002, park_wavelet_2007,preskill_quantum_2018}. Among the various physical platforms for qubit implementation, superconducting systems are prominent candidates due to their rapid gate operation times and robust scalability \cite{namkung_coherence_2020, choe_efficient_2025,arute_quantum_2019, kim_evidence_2023, blais_circuit_2021, krantz_quantum_2019}. Transitioning from the Noisy Intermediate-Scale Quantum (NISQ) era to practical quantum computation necessitates large-scale Quantum Error Correction (QEC) \cite{ kjaergaard_superconducting_2020}.

Surface codes, which implement logical qubits on a lattice structure, are central to QEC and are the subject of intense research within superconducting hardware development \cite{fowler_surface_2012, wang_surface_2011, google_quantum_ai_suppressing_2023, google_quantum_ai_and_collaborators_quantum_2025, vezvaee_surface_2025, kim_design_2023, kim_implementation_2025, kim_about_2025}. To date, superconducting architectures have primarily diverged into grid-based layouts (Google) and heavy-hexagon structures (IBM). While Google's grid architecture—incorporating individual couplers between qubit pairs—allows for direct surface-code mapping, the escalating number of couplers in larger systems triggers performance degradation through unwanted interactions and crosstalk \cite{ding_systematic_2020}. Conversely, IBM's heavy-hexagon structure suppresses these interactions by reducing connectivity \cite{chamberland_topological_2020}, yet it incurs mapping inefficiencies, such as the requirement for additional SWAP gate overhead. Furthermore, many current platforms utilize tunable couplers for gate control \cite{yan_tunable_2018, sete_floating_2021}, 
which require dedicated flux-bias lines and expose the system to decoherence from flux noise \cite{paladino_1_2014}.

To address these structural bottlenecks, research has shifted toward alternative hardware architectures. Recent proposals have explored coupling three fixed-frequency qubits to a single fixed-frequency coupler \cite{kang_new_2025}, demonstrating high gate fidelities via microwave-based control \cite{chow_simple_2011, shirai_all-microwave_2023, jiang_microwave-activated_2025}. However, these configurations still face structural constraints when aligning with surface-code lattice units, necessitating complex mapping procedures for large-scale expansion.

In this work, we propose a "lattice-patch" architecture that integrates four fixed-frequency transmon qubits with a single fixed-frequency coupler. This design achieves topological alignment with the surface code, overcoming the mapping limitations of previous structures. By employing an entirely fixed-frequency configuration, we minimize sensitivity to flux noise and maximize physical stability. Through numerical simulations, we demonstrate CNOT gates across six distinct directions within the patch, establishing this architecture as an efficient and stable building block for next-generation fault-tolerant quantum computers.

\section{System Configuration} \label{sec:system}

The proposed architecture integrates four fixed-frequency transmon qubits with a single fixed-frequency coupler, a configuration we define as the Lattice-Transmon system.

\subsection{Comparison of Hardware Architecture and Connectivity}

\begin{figure*}[htbp]
  \centering
  \subfloat[\label{fig:sub2}]{\includegraphics[width=0.6\columnwidth]{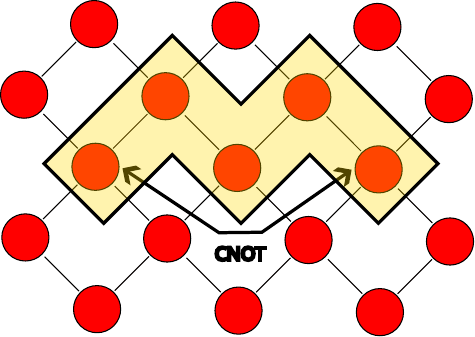}}
  \hfill 
  \subfloat[\label{fig:sub2}]{\includegraphics[width=0.6\columnwidth]{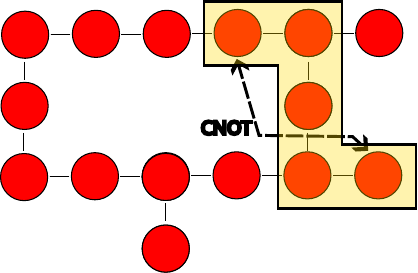}}%
  \hfill
  \subfloat[\label{fig:sub3}]{\includegraphics[width=0.8\columnwidth]{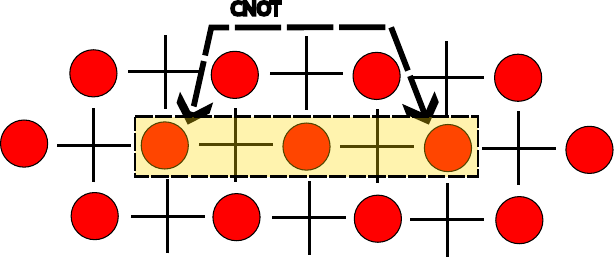}}
  
  \caption{Comparison of prominent superconducting quantum computing architectures. (\textbf{a}) Google's grid architecture utilizes individual couplers between adjacent qubits, enabling direct surface-code mapping; however, the high coupler density in scaled systems leads to performance degradation from parasitic interactions and crosstalk. (\textbf{b}) IBM's heavy-hexagon structure suppresses these interactions by reducing qubit connectivity, yet it lacks direct surface-code mapping, resulting in reduced mapping efficiency and significant SWAP gate overhead. (\textbf{c}) The proposed lattice-patch architecture couples four fixed-frequency transmons to a single fixed-frequency coupler, facilitating direct topological correspondence with surface-code lattice units [plaquettes] while providing high connectivity without additional hardware overhead. \label{fig:system_structure}}
\end{figure*}

Figure \ref{fig:system_structure} illustrates the structural advantages of the Lattice-Transmon system relative to representative superconducting quantum computing architectures.

Conventional designs, such as Google's grid architecture (Fig. \ref{fig:system_structure}a) and IBM's heavy-hexagon layout (Fig. \ref{fig:system_structure}b), deploy individual couplers between every adjacent qubit pair to mediate direct interactions \cite{arute_quantum_2019, kim_evidence_2023}. As systems scale, this approach leads to a prohibitive increase in the number of couplers and control lines. Moreover, during $CNOT$ operations on a target qubit pair, these architectures are highly susceptible to parasitic interactions from the dense surrounding network of couplers and qubits \cite{ding_systematic_2020}. 
In contrast, our proposed architecture (Fig. \ref{fig:system_structure}c) utilizes a patch-based geometry where four qubits are locally coupled through a single central mediator. By minimizing the number of physical components involved in gate operations, this design significantly reduces hardware complexity and mitigates operational overhead \cite{wu_modular_2024}.

\subsection{Physical Model and Energy Level Design}

\begin{figure*}[htbp]
  \centering
  \subfloat[\label{fig:sub1}]{\includegraphics[width=0.5\columnwidth]{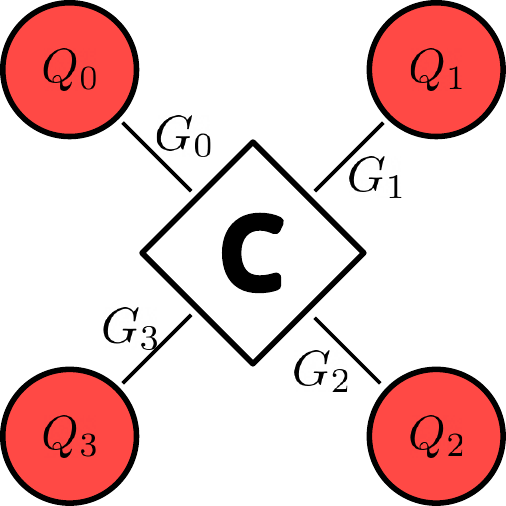}}%
  \hfill 
  \subfloat[\label{fig:sub2}]{\includegraphics[width=0.7\columnwidth]{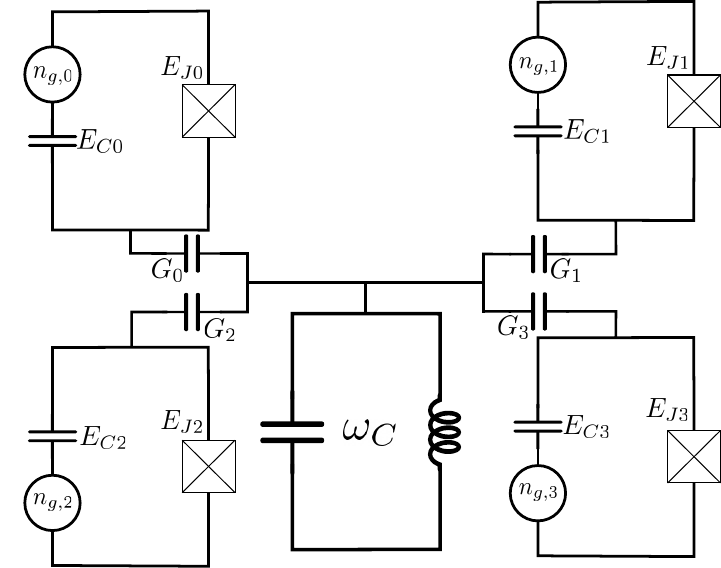}}%
  \hfill
  \subfloat[\label{fig:sub3}]{\includegraphics[width=0.55\columnwidth]{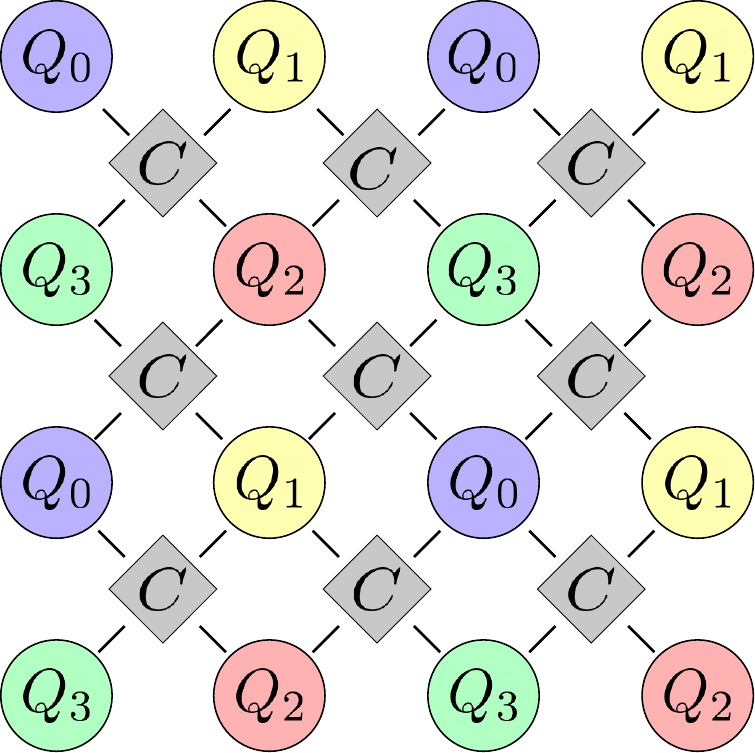}}
  
  \caption{
	Detailed architecture and circuit model of the lattice-patch structure. (\textbf{a}) Schematic of the 1-coupler-4-qubit system. The Lattice-Transmon architecture consists of four fixed-frequency transmon qubits ($Q_{0}$-$Q_{3}$) coupled to a single fixed-frequency central mediator (C). Each qubit is locally connected via coupling strengths $G_{i}$, a configuration designed to minimize gate operation overhead. (\textbf{b}) Lumped-element circuit model. The central coupler is implemented as an LC resonator with resonance frequency $\omega_{C}$. Each transmon is defined by its Josephson energy ($E_{J,i}$) and charging energy ($E_{C,i}$). External microwave drive voltages ($n_{g,i}$) facilitate qubit control, while capacitive coupling mediates interactions between the qubits and the central resonator. (\textbf{c}) Scalability of the lattice patch. The periodic tiling of the patch unit drastically reduces the total coupler count compared to standard architectures and enables direct mapping onto the surface-code lattice.
	\label{fig:lattice_transmon}}
\end{figure*}

The detailed architecture and circuit model of the proposed system are illustrated in Fig. \ref{fig:lattice_transmon}. The central fixed-frequency coupler is designed as an LC resonator, to which four transmon qubits are capacitively coupled with strengths $G_{i}$ (Fig. \ref{fig:lattice_transmon}a, b).

By utilizing entirely fixed-frequency components, we ensure the system is intrinsically robust against $1/f$ flux noise—a primary decoherence mechanism in superconducting qubits. To prevent frequency crowding, the operating frequencies of the qubits ($\omega_i$) are designed with substantial detunings on the order of several hundred MHz. Furthermore, the coupler frequency is positioned approximately 2 GHz above the qubit frequency band. This high-frequency placement suppresses direct energy exchange between qubits and the coupler, allowing the latter to serve as a stable, dispersive mediator.

\subsection{Definition of the System Hamiltonian}

We formulate the system Hamiltonian $\hat{\mathcal{H}}$ using a lumped-element circuit model. For computational convenience, we set $\hbar=1$. The total Hamiltonian is defined as the sum of the resonator ($\hat{H}_{R}$), transmon ($\hat{H}_{T}$), and interaction ($\hat{H}_{I}$) terms:
\begin{equation}
    \hat{H}=\hat{H}_{R}+\sum_{i=0}^{3}\hat{H}_{T,i}(t)+\hat{H}_{I},
\end{equation}

First, the coupler, implemented as an LC circuit, is modeled as a quantum harmonic oscillator. The Hamiltonian for the resonance frequency $\omega_{C}$ is given by:
\begin{equation}
    \hat{H}_{R}=\omega_{C}\hat{a}^{\dagger}\hat{a},
\end{equation}
where $\hat{a}^{\dagger}$ and $\hat{a}$ denote the creation and annihilation operators, respectively.

Second, the Hamiltonian for each transmon qubit, $\hat{H}_{T,i}$, is defined by its charging energy ($E_{C,i}$) and Josephson energy ($E_{J,i}$) \cite{kochChargeinsensitiveQubitDesign2007, nakamura_coherent_1999}:
\begin{equation}
    \hat{H}_{T,i}(t)=4E_{C,i}(\hat{n}_{i}-n_{g,i}(t))^{2}-E_{J,i}\cos\hat{\phi}_{i},
\end{equation}
where $\hat{n}{i}$ is the Cooper-pair number operator, $\hat{\phi}_{i}$ is the phase operator, and $n_{g,i}(t)$ represents the external microwave voltage pulses applied for qubit control. The system is designed in the transmon regime ($E_{J} \gg E_{C}$) to ensure insensitivity to charge noise.

Third, the interaction between the coupler and qubits follows the Jaynes-Cummings model and is expressed as \cite{shoreJaynesCummingsModel1993}:

\begin{equation}
    \hat{H}_{I}= \sum_{i=0}^{3}G_{i}(\hat{a}+\hat{a}^{\dagger})\hat{n}_{i}.
\end{equation}
where $G{i}$ represents the coupling strength between the coupler and the $i$-th qubit.

Given the reliance on fixed-frequency elements, minimizing the residual $ZZ$ interaction during the idle state is critical \cite{ni_scalable_2022}. By optimizing qubit detunings within the dispersive regime, we suppress unintended parasitic phase accumulation in the computational basis. This design strategy serves as a foundation for mitigating crosstalk in large-scale lattice expansions.

\subsection{Numerical Simulation and Multi-level Considerations}

To accurately simulate gate dynamics within a realistic hardware environment, we model the transmons as multi-level systems rather than simple two-level systems. This approach enables a rigorous analysis of leakage errors into non-computational states during gate execution \cite{aliferis_fault-tolerant_2007}. The total Hilbert space of the system is defined by the tensor product of the resonator's Fock states and the eigenbasis of each transmon:

\begin{equation}
    \mathcal{S}=\vert k\rangle_{R}\otimes \bigotimes_{i=0}^{3}\vert m_{i}\rangle ,\quad(k,m_{i}\in {0,1,2}).
\end{equation}
To balance computational efficiency with numerical accuracy, we include the three lowest energy levels ($|0\rangle, |1\rangle, |2\rangle$) for each component. By monitoring state transitions out of the computational basis $\{|0\rangle, |1\rangle\}$, we identified the optimal parameters required for achieving high-fidelity gate performance.

\section{Pulse protocol and Quantum Gatme Optimization} \label{sec:pulse}
In this section, we discuss the pulse design strategies and numerical optimization procedures required to implement high-fidelity two-qubit gates within the constrained lattice-patch architecture. Our control protocol utilizes a cross-resonance (CR) pulse-based framework, which has been previously validated in a system coupling three fixed-frequency qubits to a single fixed-frequency coupler. \cite{magesan_effective_2020} By adapting and expanding this established approach to our proposed architecture , we aim to demonstrate that structural scalability and operational consistency can be maintained even under a more complex interaction environment.

\subsection{Control Strategy and Scability}
Compared to the previous researching structure coupling three qubits to a single fixed-frequency coupler, the proposed architecture introduces a more pronounced frequency crowding problem, which inherently exacerbates parasitic crosstalk. Despite these physical challenges, we retain the exact pulse framework established in previous literature to verify the physical viability and operational feasibility of our proposed hardware. Maintaining this identical pulse framework ensures its universal applicability within the local patch , thereby offering significant architectural advantages for scaling up large-scale quantum processors.

\subsection{Parameter Optimization}
Optimizing pulse parameters in superconducting qubit systems not only demonstrates hardware viability but also elucidates the potential limitations and future scalability of the proposed system. To implement the system, we employ two primary pulse types: a CR pulse to induce conditional interactions between the control and target qubits, and a DRAG-based Gaussian flat-top pulse to finalize the CNOT gate implementation. Optimizing the gates using each pulse type serves as a critical metric for evaluating the performance of the proposed system, making the optimization process a key technical challenge. To optimize these pulse parameters, we performed a grid search via a comprehensive parameter sweep.

In this simulation, to demonstrate that CNOT gates can be implemented across all connectivity directions , we realized a total of twelve gates: six forward CNOT gates ($CNOT_{01},CNOT_{02},CNOT_{03},CNOT_{12},\\ CNOT_{13},CNOT_{23}$) and six reverse CNOT gates ($CNOT_{10},CNOT_{20},CNOT_{21},CNOT_{30},CNOT_{31}, \\ CNOT_{32}$). For the reverse gates, we leveraged the pulses already developed for the forward operations and applied an additional Gaussian pulse to achieve the reverse implementation. Specifically, the pulse used for the reverse operations was configured by additionally applying the same Gaussian flat-top gate sequence used in the forward CNOT gate framework. Consequently, to optimize the reverse CNOT gates, it is also necessary to optimize the auxiliary parameters introduced by these additional pulses.

\subsection{Pulse Parameter Initization and Pulse composition}
To implement the CNOT gate, the cross-resonance (CR) pulse is characterized by five parameters,  $(f_{1}, T_{S},\Omega_{S},\rho,\gamma_{1})$, while the Gaussian pulse relies on four parameters, $(f_{2},T_{X},\Omega_{X},\gamma_{2})$. Consistent with the approach used in the 1-coupler-3-qubit system, the initial values for the CR pulse parameters were set to $(f_{1}, T_{S},\Omega_{S},\rho,\gamma_{1})=(f_{\text{target qubit}},150, 0.05,0.1,0)$. A grid search was subsequently conducted via parameter sweeps with step sizes or ranges of $[0.01,5,0.01,5,0]$ for each respective parameter.

For the Gaussian pulse, the four parameters $(f_{2},T_{X},\Omega_{X},\gamma_{2})$were initialized to $(f_{2},T_{X},\Omega_{X},\gamma_{2})=(f_{\text{target qubit}}10,0.01,0)$, following the same methodology as the 1-coupler-3-qubit system. This was also optimized using a grid search via parameter sweeps within the bounds of $[0.01,1,0.005, 0.01\pi]$. In particular, the phase parameter for the Gaussian pulse was meticulously scanned across the range of $[-\pi,\pi]$ with a fine interval of 0.01.

It is worth noting that in demonstrating the reverse gate fidelities in this study, the pulses for single-qubit gates and their associated physical errors were not explicitly modeled. This is because the single-qubit error rate in current superconducting transmon processor platforms is remarkably low $(\sim 10^{-4})$, making it negligible compared to that of two-qubit gates like the CNOT gate. Furthermore, the primary objective of this work is to introduce and validate a novel architecture where four qubits are coupled to a single mediator. Specifically, the implementation of the reverse gate aims to confirm that a bidirectional CNOT gate can be successfully achieved not by driving the conventional control qubit, but rather by applying the CR pulse to the target qubit.

\begin{figure*}[htbp]
  \centering
  \subfloat[\label{fig:sub1}]{\includegraphics[width=1\columnwidth]{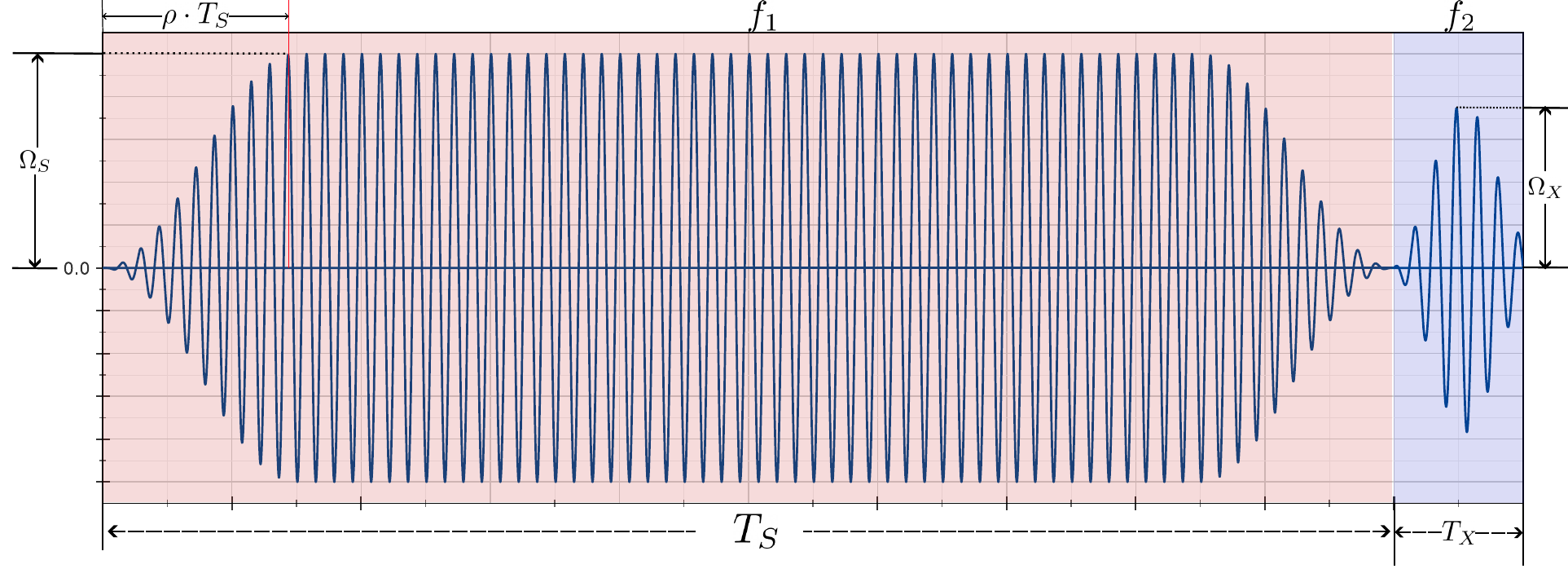}}%
  \hfill 
  \subfloat[\label{fig:sub2}]{\includegraphics[width=1\columnwidth]{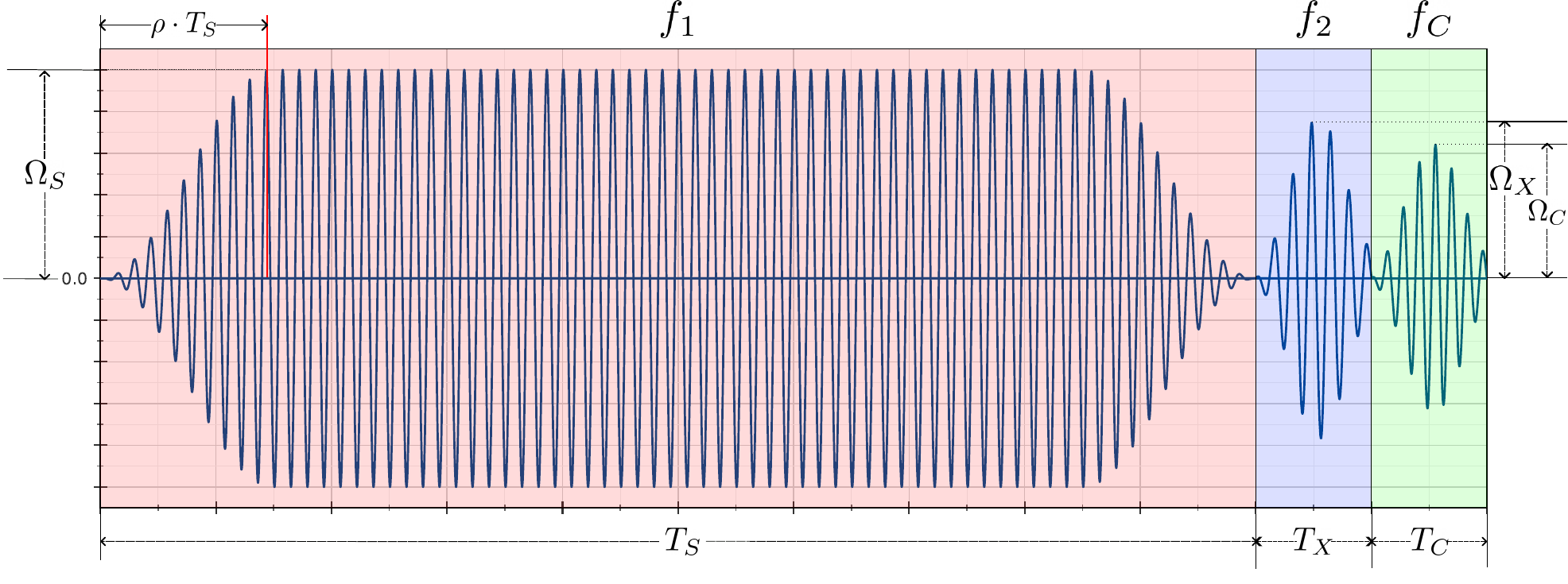}}%

  \caption{
	Pulse waveforms for gate implementation. (\textbf{a}) Pulse sequences for the forward CNOT gate. The red and blue regions represent the drives applied to the control and target qubits, respectively. The control qubit is driven by a cross-resonance (CR) pulse, where $T_{S}$, $f_{1}$, $\Omega_{S}$, $\gamma$, and $\rho$ denote the pulse duration, frequency, amplitude, initial phase, and ramp ratio, respectively; the ramp ratio is scaled relative to the total pulse duration. The target qubit drive consists of a Gaussian pulse with DRAG [Derivative Removal by Adiabatic Gate] correction, characterized by duration $T_{X}$, frequency $f_{2}$, amplitude $\Omega_{X}$, and initial phase $\gamma$. (\textbf{b}) Pulse sequences for the reverse CNOT gate. The red and blue regions follow the same implementation as the forward gate, while the green region denotes an auxiliary drive applied to the control qubit specifically to enhance the fidelity of the reverse operation. \label{fig:pulse_shapes}
}
\end{figure*}

The applied microwave drive pulses $n_{g,i}(t)$ conform to the assumed physical model and are designed to interact with the four-qubit energy spectrum via the pre-configured driving Hamiltonian. Figure \ref{fig:pulse_shapes}a illustrates the pulse sequence for the forward CNOT gate, while Figure \ref{fig:pulse_shapes}b displays the pulse sequence for the reverse CNOT gate.

The pulse sequence employed here is identical to the framework used in the conventional 1-coupler-3-qubit system, and its detailed description is as follows:
\begin{enumerate}
    \item \textbf{Simple Cross-Resonance Pulse Design:} We utilized the most foundational, simple CR pulse scheme to validate the structural feasibility of our 1-coupler-4-qubit architecture. By employing this basic CR drive, our primary objective was to demonstrate that conditional dynamics between two qubits can be successfully realized even within this expanded hardware topology. 
    \item \textbf{Leakage Suppression via DRAG:} To mitigate frequency crowding and prevent leakage into higher energy levels, we apply the Derivative Removal by Adiabatic Gate (DRAG) technique to all Gaussian pulse envelopes:
    \begin{equation}
        n_{g,i}(t)=\Omega_{G}(t)\cos(2\pi f_{i}t -\gamma_{i})+\beta_{i}\dot{\Omega}_{G}\sin(2\pi f_{i}t -\gamma_{i}).
    \end{equation}
    where the scaling parameter $\beta_{i}$ is set to $0.5/E_{C}$ to effectively suppress charge-noise sensitivity and transitions to higher-order levels.
\end{enumerate}

\subsection{Flexibility in reverse-directional Gate Implementation}
Despite the reliance on fixed-frequency components, our architecture enables the efficient implementation of bidirectional CNOT gates through pulse-level calibration (Fig. \ref{fig:pulse_shapes}b). Rather than wrapping a forward-direction gate with virtual Hadamard transformations, we ensure gate robustness by physically optimizing auxiliary pulses. This optimization explicitly accounts for the hardware-specific coupling strengths ($G_{i}$) and frequency detunings, allowing for high-performance operations in both directions.

\subsection{Numerical Optimization and Validation Methodology}
To evaluate the impact of the designed pulses on the system Hamiltonian and derive the final unitary operator $\hat{\mathcal{U}}$, we perform numerical optimization based on the time-dependent Schrödinger equation. The system Hamiltonian is formulated as the sum of a static term ($\hat{H}_{0}$) and a pulse-controlled driving term ($\hat{H}_{1}(t)$):

\begin{equation}
    \hat{H}(t)=\hat{H}_{0}+\hat{H}_{1}(t).
\end{equation}

We calculate the time evolution over the total pulse duration $T$ using a Dyson series expansion. For enhanced numerical precision, we employ a Suzuki-Trotter decomposition \cite{argeri_magnus_2014, schmitz_graph_2024}:
\begin{widetext}
\begin{equation}
    \exp(-i \tau \hat{H}(\tilde{t})) \simeq \exp(-i \tau \hat{H}_{0}/2) V \exp(-i \tau \hat{\Lambda}(\tilde{t}_n)) V^{\dagger} \exp(-i \tau \hat{H}_{0}/2).
\end{equation}
\end{widetext}

where $\hat{\Lambda}(\tilde{t}_n) = V^{\dagger} \hat{H}_{1}(\tilde{t}_n) V$, and $V$ denotes the transformation matrix composed of the eigenvectors corresponding to the transmon qubit states. The time step $\tau$ is chosen to guarantee numerical convergence and satisfy the required error bounds. In this study, we adopt a fine interval of $\tau=0.001$ ns to minimize higher-order truncation errors and ensure the reliability of the calculated gate fidelities.

When performing the time evolution of the aforementioned Hamiltonian, the high-frequency oscillations intrinsic to the qubits impose physical and numerical limitations on interpreting the qubit system. To circumvent this challenge, it is necessary to transition from the current laboratory (Lab) frame to a rotating frame. Since this frame transformation must be applied universally across the Hamiltonian terms partitioned via the Suzuki-Trotter decomposition, the frame transition is executed by incorporating additional gauge-like operational terms into the Hamiltonian. The corresponding governing equation can be expressed as follows:

\begin{widetext}
\begin{equation}  
  \begin{gathered}
  \exp(-i\tau\hat{H}_{0}/2) \rightarrow \exp\bigl(-i \tau (\hat{H}_{0}+i \dot{R}R^{\dagger}/2)\bigr), \\
  \hat{U}_{\text{final}} = \exp\bigl(-i \tau (\hat{H}_{0}+i \dot{R}R^{\dagger}/2)\bigr) R V \exp\bigl(-i \hat{\Lambda}(\tilde{t}_{n})\bigr) V^{\dagger} R^{\dagger} \exp\bigl(-i \tau (\hat{H}_{0}+i \dot{R}R^{\dagger}/2)\bigr).
  \end{gathered}
\end{equation}
\end{widetext}


where $R$ and $\dot{R}$ are defined as follows:
\begin{widetext}
\begin{equation}
	R(t)= \begin{bmatrix}
0 & 0 & 0\\
0 & e^{i\omega t} & 0 \\
0 & 0 & e^{2i\omega t}
	\end{bmatrix},\qquad \dot{R}=\frac{d}{dt}R(t)= \begin{bmatrix}
0 & 0 & 0\\
0 & i\omega e^{i\omega t} & 0 \\
0 & 0 & 2i\omega e^{2i\omega t}.
	\end{bmatrix}
\end{equation}
\end{widetext}

By incorporating these frame-transformation terms, we perform numerical simulations that effectively account for the high-frequency dynamics within the Hamiltonian evaluation.

Furthermore, we apply virtual-Z gate techniques to compensate for residual phase errors post-pulse application, yielding the final unitary matrix $\hat{\mathcal{U}}{\text{total}}$ \cite{ghosh_high-fidelity_2013}:
\begin{equation}
    \hat{R}_{z}^{(\mathbf{v})}(\theta)=\bigotimes_{i=0}^{3}R_{z}^{(i)}(\theta_{i}),\quad \hat{\mathcal{U}}_{\text{total}}=\hat{R}_{z}^{(\mathbf{v})}\hat{\mathcal{U}}.
\end{equation}

Gate performance is quantified by the average gate fidelity ($F_{\text{avg}}$), calculated via numerical Monte Carlo integration over $M=10^{4}$ samples:
\begin{equation}
    F_{\text{avg}}=\frac{1}{M}\sum_{j=1}^{M} |\langle\psi_{j}\vert \hat{\mathcal{U}}_{\text{opt}}^{\dagger} \hat{\mathcal{U}}_{\text{total}}\vert\psi_{j} \rangle|^{2}.
\end{equation}

Since the final unitary gate incorporated with the virtual-Z (VZ) gates may not yet represent a fully optimized pulse configuration, an additional fine-tuning optimization process is conducted. Based on the approximate parameter regions identified via the initial grid search sweep, these discovered points are set as the initial coordinates. Subsequently, the derivative-free Nelder-Mead simplex algorithm is executed to further enhance the fidelity and performance of the implemented gate. The comprehensive optimization protocol for the final unitary operator $\mathcal{U}_{\text{final}}$ is structured as follows:

\begin{enumerate}
    \item \textbf{Initial Point Setup:} \\
    The parameters swept during the primary CR and Gaussian pulse calibrations are utilized to construct the initial points within the proximity of the optimal region. Starting the derivative-free Nelder-Mead algorithm from these pre-localized coordinates is crucial to prevent the optimizer from getting trapped in unfavorable local minima.
    
    \item \textbf{Loss Function Setup:} \\
    To drive the optimization process, we define a loss function $\mathcal{L}$. This loss is formulated using the average gate fidelity $F_{\text{avg}}$ calculated from the implemented unitary operator $\mathcal{U}_{\text{total}}$ evaluated over random states $\vert \psi \rangle$ spanning the computational basis. The loss function $\mathcal{L}$ is computed as follows:
    \begin{equation}
        \mathcal{L} = 1 - F_{\text{avg}}.
    \end{equation}
    
    \item \textbf{Convergence Criteria:} \\
    The optimization process is terminated and deemed converged when either the variation in the simplex size $\Delta x$ within the parameter space or the fractional change in the loss function $\Delta \mathcal{L}$ drops below a predefined threshold of $10^{-4}$, indicating that no further numerical improvement is achievable.
\end{enumerate}

This rigorous validation framework demonstrates that the proposed lattice-patch architecture achieves high operational reliability within a complex four-qubit network while maintaining the control simplicity of previous microwave-only schemes.

\section{Setting Parameters of the Lattice-Transmon System} \label{sec:results}

In this section, we present the hardware specifications and optimized gate parameters to demonstrate the physical viability of the proposed Lattice-Transmon system. We further validate the architectural effectiveness through numerical simulations, analyzing the resulting success probabilities and gate fidelities.

\subsection{Hardware Specifications and Frequency Allocation}

The hardware parameters for the proposed system are detailed in Table \ref{tab:hardwareparameter}. This study extends a previously validated model to a four-qubit architecture, with qubit operating frequencies ranging from $\omega_{0}/2\pi = 5.03260$ GHz to $\omega_{3}/2\pi = 4.67259$ GHz. These frequencies are designed to coincide with the eigenfrequencies of the system Hamiltonian in the absence of external driving pulses.

\begin{table}[b] 
\caption{Hardware specifications. \label{tab:hardwareparameter}}
\begin{ruledtabular} 
\begin{tabular}{cccccc} 
  & C & \textbf{$Q_{0}$}  & \textbf{$Q_{1}$} & \textbf{$Q_{2}$} & \textbf{$Q_{3}$} \\
\hline 
$\omega/2\pi$   & 7.0 & 5.03260 & 4.91339 & 4.79421 & 4.67259 \\
$E_{C,i}/2\pi$  & --  & 0.33    & 0.32    & 0.31    & 0.3     \\
$E_{J,i}/2\pi$  & --  & 11.0    & 10.8    & 10.6    & 10.4    \\
$G_{i}/2\pi$    & --  & 0.06    & 0.05    & 0.03    & 0.06    \\ 
\end{tabular} 
\end{ruledtabular}
\end{table}

To mitigate frequency crowding—a common bottleneck in large-scale processors—we maintained inter-qubit detunings between 100 and 300 MHz. Furthermore, by assigning distinct qubit-coupler coupling strengths ($G_{i}$), we analyzed how variations in coupling affect gate implementation and overall performance. To quantify the crosstalk in our system, the residual $ZZ$ interaction strength $\zeta_{ij}$ between qubits $i$ and $j$ is defined as follows:

\begin{equation}
	\zeta_{ij}=E_{\vert 1_{i}1_{j}\rangle}-E_{\vert 1_{i}0_{j}\rangle}-E_{\vert 0_{j}1_{i}\rangle}+E_{\vert 0_{j}0_{i}\rangle}.
\end{equation}
Moreover, to evaluate the multi-body crosstalk across the entire four-qubit system, we investigate the fourth-order residual $ZZ$ interaction $\zeta_{1111}$, which is expressed as:

\begin{widetext}
\begin{equation}
	\begin{aligned}
	\zeta_{1111} =& E_{\vert 1110\rangle}-(E_{\vert 1101\rangle} + E_{\vert 1011\rangle} + E_{\vert 0111\rangle} )\\
	&+(E_{\vert 1100\rangle}+E_{\vert 1010\rangle} + E_{\vert 1001\rangle} + E_{\vert 0110\rangle} + E_{\vert 0101\rangle} + E_{\vert 0011\rangle})\\
	&-(E_{\vert 0001\rangle} +E_{\vert 0010\rangle} + E_{\vert 0100\rangle} +E_{\vert 1000\rangle}) + E_{\vert 1111\rangle}
	\end{aligned}
\end{equation}
\end{widetext}

The resulting residual $ZZ$ interactions for the system are summarized in Table \ref{tab:zz_interaction}.

\begin{table*}[t] 
\caption{Residual $ZZ$ Interactions (MHz).\label{tab:zz_interaction}}
\begin{ruledtabular}
\begin{tabular}{ccccccc}
  $\zeta_{01}$ & $\zeta_{02}$ & $\zeta_{03}$ & $\zeta_{12}$ & $\zeta_{13}$ & $\zeta_{23}$ & $\zeta_{1111}$  \\
\hline
  0.02339  & 0.01164 & 0.16352 & 0.00536  & 0.03352 & 0.00811 & 0.23288 \\
\end{tabular}
\end{ruledtabular}
\end{table*}

\subsection{Optimization of CNOT Gate Pulse Parameters}

Table \ref{tab:pulse_paramter} lists the optimized pulse parameters for the six forward-direction CNOT gates implemented in this architecture. Each gate protocol was optimized by applying a Cross-Resonance (CR) drive at the target qubit frequency ($f_{2}$), integrated with Gaussian auxiliary pulses to cancel parasitic interactions. To maximize gate fidelity, we performed fine-frequency sweeps within a $\pm 10$ MHz range and optimized the ramp parameter ($\rho$) to refine the pulse rise and fall times, thereby minimizing leakage into higher-order energy levels.

\begin{table*}[t] 
\caption{Forward CNOT gate parameters\label{tab:pulse_paramter}}
\begin{ruledtabular}
\begin{tabular}{cccccccccc}
\textbf{Gate} & \textbf{$f_{1}$} [GHz] & \textbf{$f_{2}$} [GHz] & \textbf{$T_{S}$} [ns] & \textbf{$T_{X}$} [ns] & \textbf{$\Omega_{S}$} & \textbf{$\Omega_{X}$} & \textbf{$\rho$} & \textbf{$\gamma_{1}$} & \textbf{$\gamma_{2}$}\\
\hline
$\text{CNOT}_{01}$ & 4.91339 & 4.91339 & 220  & 9 & 0.05 & 0.01  & 0.178 & 0 & 0.776 \\
$\text{CNOT}_{02}$ & 4.79396 & 4.79421 & 295  & 10 & 0.07 & 0.01  & 0.12 &0 & 0.5\\
$\text{CNOT}_{03}$ & 4.67257 & 4.67259 & 200  & 10 & 0.06 & 0.025 & 0.3 &0 & -0.05 \\
$\text{CNOT}_{12}$ & 4.79398 & 4.79421 & 440  & 10 & 0.05  & 0.017 & 0.1 &0 & 0.03 \\
$\text{CNOT}_{23}$ & 4.67298 & 4.67259 & 400  & 10 & 0.075  & 0.003 & 0.1  &0 &0.87 \\
\end{tabular}
\end{ruledtabular}
\end{table*}

The auxiliary drives used for these reverse operations share the same Gaussian flat-top envelope configuration as the forward gates. Specifically, the optimized pulse parameters are specified by the tuple $(f, T, \Omega, \gamma)$, which represent the target qubit frequency, total pulse duration, dimensionless pulse amplitude, and initial phase, respectively. Detailed parameters for each reverse CNOT gate configuration are provided in Table \ref{tab:pulseparameter_inv} and Table \ref{tab:remained_cnot_gate}.

\begin{table}[t] 
\caption{Forward CNOT gate parameters\label{tab:pulseparameter_inv}}
\begin{ruledtabular}
\begin{tabular}{ccccc}
\textbf{Gate} & \textbf{$f$} [GHz] & \textbf{$T$} [ns] & \textbf{$\Omega$} & \textbf{$\gamma$} \\
\hline
$\text{CNOT}_{10}$ & 4.91339 &  16 & 0.03 & -0.78 \\
$\text{CNOT}_{20}$ & 4.79421 & 45 & 0.04  & 0.5 \\
$\text{CNOT}_{30}$ & 4.67259 & 42 & 0.01   & -0.15 \\
$\text{CNOT}_{21}$ & 4.99421 & 41 & 0.032   & -0.9 \\
$\text{CNOT}_{32}$ & 4.67259 & 10 & 0.015 & 0.6 \\ 
\end{tabular}
\end{ruledtabular}
\end{table}

However, the $\text{CNOT}_{31}$ gate exhibited a trend where fidelity failed to improve despite the application of ideal Hadamard gates and subsequent pulse optimization. Consequently, we configured the $\text{CNOT}_{31}$ gate using a direct forward-drive pulse sequence rather than the reverse-wrapping method, as detailed in Table \ref{tab:remained_cnot_gate}.

\begin{table*}[t] 
\caption{CNOT13, CNOT31 pulse parameter\label{tab:remained_cnot_gate}}
\begin{ruledtabular}
\begin{tabular}{cccccccccc}
\textbf{Gate} & \textbf{$f_{1}$} [GHz] & \textbf{$f_{2}$} [GHz] & \textbf{$T_{S}$} [ns] & \textbf{$T_{X}$} [ns] & \textbf{$\Omega_{S}$} & \textbf{$\Omega_{X}$} & \textbf{$\rho$} & \textbf{$\gamma_{1}$} & \textbf{$\gamma_{2}$}\\
\hline
$\text{CNOT}_{13}$ & 4.67259 & 4.91339 & 210  & 10 & 0.085 & 0.02  & 0.17 & 0 & 0.75398 \\
$\text{CNOT}_{31}$ & 4.91369 & 4.67259 & 430  & 20 & 0.09 & 0.0227 & 0.1 & 0 & -1.13097 \\
\end{tabular}
\end{ruledtabular}
\end{table*}

This observation suggests that increasing the qubit-to-coupler ratio narrows the frequency gaps between qubits, exacerbating frequency crowding. These results underscore the critical importance of frequency allocation strategies in large-scale systems, highlighting the need for sophisticated frequency-placement and dynamic tuning techniques in future research.

\subsection{Analysis of Gate Fidelity and Success Probability}
The final unitary operator is implemented by sequentially applying the virtual-Z (VZ) gates immediately following the primary pulse operations. Table \ref{tab:apply_correcting_vz} lists the precise rotation angles ($\theta$) applied to each individual qubit to calibrate the pulse-induced unitary operator.

To evaluate the performance of the implemented gates, we performed unitary operations on prepared computational basis states and calculated the resulting success probabilities. As shown in Table \ref{tab:final_resutls} , the forward CNOT gates ($\text{CNOT}_{01}$ through $\text{CNOT}_{23}$) exhibit high success probabilities exceeding 0.98. High success rates were maintained across varying coupling strengths, albeit with longer gate durations, suggesting that the proposed lattice-patch architecture possesses high robustness against physical parameter fluctuations.

\begin{table*}[t] 
\caption{Apply correcting VZ gate\label{tab:apply_correcting_vz}}
\begin{ruledtabular}
\begin{tabular}{ccccc||ccccc}
\textbf{Gate}	& \textbf{$\theta_{0}$}	& \textbf{$\theta_{1}$}  & \textbf{$\theta_{2}$} & \textbf{$\theta_{3}$} &   \textbf{Gate} & \textbf{$\theta_{0}$}	& \textbf{$\theta_{1}$}     & \textbf{$\theta_{2}$} & \textbf{$\theta_{3}$} \\
\hline
$\text{CNOT}_{01} $ & -0.5966 & -0.0314 & 0 & 0.3142 & $\text{CNOT}_{10}$ & 0.0942 & -0.0314 & 0 & 0 \\
$\text{CNOT}_{02} $ & 0.4712 & 0 & 0 & 0 & $\text{CNOT}_{20}$ &  0 & 0.0628 & 0 & 0  \\
$\text{CNOT}_{03} $ & -0.2512 & 0 & 0 & 0.3142 & $\text{CNOT}_{30}$ & 0.0314 & 0.0314 & 0.0314 & 0.0628\\
$\text{CNOT}_{12} $ & 0.1884 & 1.4727 & 0.2826 & 0.2198 & $\text{CNOT}_{21}$ & 0.0314 & 0.1256 & -0.0314 & -0.0314  \\
$\text{CNOT}_{13} $ & 0.0942 & 0.942 & 0 & -0.3768  & $\text{CNOT}_{31}$ & 0.0942 & 0 & 0.0314 & 0.0942\\
$\text{CNOT}_{23} $ & 0.1727 & 0.0942 & -0.0628 & 0.1099 &$\text{CNOT}_{32}$ & 0.1256 & 0.0942 & 0.157 & 0.1256  \\
\end{tabular}
\end{ruledtabular}
\end{table*}

The final average gate fidelities, calculated after virtual-Z phase corrections, are summarized in Table \ref{tab:final_resutls}. The calibrated unitary gates achieved the target performance metrics across all connectivity directions. These numerical results demonstrate that our 1-coupler-4-qubit architecture can serve as an efficient building block for next-generation, large-scale quantum processors.

\begin{table*}[t] 
\caption{Lattice-Transmon system results \label{tab:final_resutls}}
\begin{ruledtabular}
\begin{tabular}{ccc||ccc}
\textbf{Gate} & \textbf{Max Success Probability} & \textbf{Average Fidelity}  & \textbf{Gate} & \textbf{Max Success Probability} & \textbf{Average Fidelity}\\
\hline
$\text{CNOT}_{01}$ & 0.9912 & 0.98265 & $\text{CNOT}_{10}$ & 0.9878 & 0.98152 \\
$\text{CNOT}_{02}$ & 0.9991 & 0.95864 & $\text{CNOT}_{20}$ & 0.9657 & 0.94496 \\
$\text{CNOT}_{03}$ & 0.9985 & 0.97833 & $\text{CNOT}_{30}$ & 0.9688 & 0.94266 \\
$\text{CNOT}_{12}$ & 0.9982 & 0.91509 & $\text{CNOT}_{21}$ & 0.9081 & 0.89566 \\
$\text{CNOT}_{13}$ & 0.9949 & 0.83292 & $\text{CNOT}_{31}$ & 0.9644 & 0.85239 \\
$\text{CNOT}_{23}$ & 0.9745 & 0.91256 & $\text{CNOT}_{32}$ & 0.9956 & 0.95000 \\
\end{tabular}
\end{ruledtabular}
\end{table*}

Here, the maximum success probability serves as a metric to evaluate how closely the implemented operator aligns with the ideal gate, evaluated across all computational basis states of the four-qubit system. Because this specific indicator is calculated by discarding phase errors, it inherently yields relatively higher numerical values. In contrast, the average gate fidelity represents the comprehensive mean value obtained only after executing the additional virtual-Z (VZ) gate calibrations. Crucially, because it encapsulates both state-flip errors and phase errors simultaneously, the average fidelity more accurately and rigorously reflects the actual gate performance from a physical hardware perspective. 

Particularly, in superconducting qubit platforms, VZ gates are executed entirely through digital frame rotations without applying any physical microwave drive pulses; thus, they introduce zero additional gate overhead or physical error in actual experimental implementations. Furthermore, the auxiliary calibration pulses employed for the reverse CNOT gate implementations were fully integrated into the time-evolution calculations of the total system Hamiltonian during the numerical simulations. Consequently, any potential physical errors induced by these secondary drives, such as parasitic crosstalk or leakage to higher energy levels, are strictly and comprehensively reflected within the final average fidelity values.

\subsection{Leakage Error and Error budget}
Because the current numerical simulations explicitly incorporate energy levels higher than the computational basis, the formulated system Hamiltonian naturally includes these higher-dimensional terms. Consequently, by applying the implemented unitary operator to the computational basis, we can quantitatively assess the leakage error. This can be formalized by projecting the final state $\rho_{\text{final}}$, which evolves from an arbitrary initial computational state $\vert \psi \rangle$ under the action of $\hat{\mathcal{U}}_{\text{final}}$, onto the respective Pauli operators $\sigma_{x}$, $\sigma_{y}$, and $\sigma_{z}$:

\begin{equation}
    x = \text{Tr}\left(\sigma_{x} \rho_{\text{final}}\right),\quad y = \text{Tr}\left(\sigma_{y} \rho_{\text{final}}\right),\quad z = \text{Tr}\left(\sigma_{z} \rho_{\text{final}}\right).
\end{equation}
where the density matrix $\rho_{\text{final}}$ is defined as follows:

\begin{equation}
    \rho_{\text{final}} = \hat{\mathcal{U}}_{\text{final}}\vert \psi \rangle \langle \psi \vert \hat{\mathcal{U}}_{\text{final}}^{\dagger}.
\end{equation}

Through this procedure, the final state evolving within the higher-dimensional Hilbert space can be effectively projected back onto the two-dimensional computational subspace. Using these calculated expectation values $(x, y, z)$, the leakage out of the computational subspace can be quantitatively measured. The resulting leakage error $\varepsilon_{L}$ is expressed by the following equation:

\begin{equation}
	\varepsilon_{L}= 1 -( x^{2}+y^{2}+z^{2}).
\end{equation}
The calculated average leakage errors for the implemented CNOT gates across the various computational basis states are summarized in the following table \ref{tab:leakage_error}.

\begin{table*}[t] 
\caption{Final CNOT gate leakage error\label{tab:leakage_error}}
\begin{ruledtabular}
\begin{tabular}{cccccccccc}
 & CNOT$_{01}$ & CNOT$_{02}$ & CNOT$_{03}$ & CNOT$_{10}$ & CNOT$_{12}$ & CNOT$_{13}$ & CNOT$_{20}$ & CNOT$_{21}$ & CNOT$_{23}$\\
\hline
$\varepsilon_{L}$ & 0.0251 & 0.0252 & 0.0238  & 0.0242 & 0.0239 & 0.0228  & 0.0244 & 0.0252 & 0.0218 \\
\end{tabular}
\end{ruledtabular}
\end{table*}

Furthermore, the physical simulations of the proposed hardware were conducted utilizing a three-level energy system ($\vert 0 \rangle, \vert 1 \rangle, \vert 2 \rangle$). This explicitly incorporates states higher than the standard two-level computational basis commonly defined in quantum computing architectures. Crucially, as the simulated energy dimension per qubit increases, the overall computational complexity of the proposed system scales as $\mathcal{O}(N^{5})$. Although the first excited state outside the computational basis (the second energy level) represents the primary channel where leakage directly occurs, higher-order leakage into even more elevated states remains a physical possibility. To investigate this phenomenon quantitatively, we tracked the specific leakage errors distributed across the higher energy levels 2, 3, and 4 during the execution of the $\text{CNOT}_{01}$ gate, the results of which are compiled in the following table \ref{tab:energy_error}.

\begin{table}[t] 
\caption{Energy Level Error Budget \label{tab:energy_error}}
\begin{ruledtabular}
\begin{tabular}{cccc}
Energy Level & 2 & 3  & 4  \\
\hline
$\varepsilon_{L}$ & 0.9941 & 0.9748 & 0.9739  \\
\end{tabular}
\end{ruledtabular}
\end{table}

On the other hand, the Hadamard gates($X_{\pi/2}$ gate) implemented for the current reverse CNOT gates were initially modeled as ideal gates. However, in actual hardware implementations, while single-qubit gates generally exhibit lower error rates compared to CNOT gates, even a minor error rate can significantly impact the overall system performance, making a detailed analysis of single-qubit gates highly critical. These single-qubit gates can be implemented using the same Gaussian pulses previously employed for the CNOT gate configuration, and the pulse parameters optimized for executing the Hadamard gates on each qubit are summarized as follows:

\begin{table}[t] 
\caption{Pulse parameters for the implemented $X_{\pi/2}$ gates. \label{tab:hadamard_gate}}
\begin{ruledtabular}
\begin{tabular}{cccccc}
\textbf{Gate} & \textbf{$f$} [GHz] & \textbf{$T$} [ns] & \textbf{$\Omega$} & \textbf{$\gamma$} & Fidelity \\
\hline
$X_{\pi/2,0}$ & 5.10260 & 10 & 0.023  & 1.5707 & 0.998 \\
$X_{\pi/2,1}$ & 4.91339 & 10 & 0.0185 &  0.3141  & 0.999 \\
$X_{\pi/2,2}$ & 4.71721 & 10 & 0.024  & -0.6283 & 0.993 \\
$X_{\pi/2,3}$ & 4.47259 & 10 & 0.025 & 0.3141  & 0.993 \\
\end{tabular}
\end{ruledtabular}
\end{table}

Accounting for these single-qubit operations, the effective fidelity of the reverse CNOT gate incorporating the physical Hadamard transformations is calculated as:
\begin{equation}
    F_{\text{eff}} \approx F_{ij} \times (1-\varepsilon_{i})(1-\varepsilon_{j})
\end{equation}
where $F_{ij}$ denotes the maximum success probability achieved by the reverse CNOT gate under idealized conditions, and $\varepsilon_{i}=1-F_{X_{\pi/2,i}}$ represents the error rate of each respective Hadamard gate. Based on this formulation, the total effective fidelities for the reverse CNOT gates are evaluated as follows:

\begin{table}[t] 
\caption{Effective fidelities of the reverse CNOT gates. \label{tab:reverse_cnot_effective_fidelity}}
\begin{ruledtabular}
\begin{tabular}{cccccc}
\textbf{Gate} & $\text{CNOT}_{10}$ & $\text{CNOT}_{20}$ & $\text{CNOT}_{21}$ & $\text{CNOT}_{30}$  & $\text{CNOT}_{32}$\\
\hline
$F_{\text{eff}}$ & 0.9848 & 0.9364 & 0.9008  & 0.9600 & 0.9817\\
\end{tabular}
\end{ruledtabular}
\end{table}

These comprehensive results demonstrate the profound impact of single-qubit errors originating from actual physical pulse sequences compared to idealized 2-qubit operations that only incorporate virtual phase corrections. Consequently, this analysis highlights that evaluating the full operation sequence is indispensable, thereby providing a rigorous and clear benchmark for assessing the practical lower bound of physical hardware performance.

\subsection{Scalability and Structural Advantages}
From a hardware implementation perspective, the 1:4 lattice-patch structure drastically reduces the total coupler count compared to conventional 1:1 or 1:2 coupling schemes, thereby mitigating control-line complexity. Furthermore, each patch functions as a unit cell [plaquette] for the surface code, and periodic tiling of these patches can form a global square lattice. This structural advantage addresses frequency crowding in fixed-frequency systems while providing an efficient physical-layer foundation for fault-tolerant quantum computing.

\section{Discussion}

\indent The single-coupler 4-qubit lattice-patch architecture proposed in this study presents a novel framework for overcoming the physical and structural constraints inherent in conventional fixed-frequency transmon systems. Our numerical simulations demonstrate the successful implementation of six independent forward CNOT gates mediated by a single LC resonator, with the flexibility to extend these operations to bidirectional gates. Despite the increased qubit density, the system achieves an average gate fidelity ($F_{\text{avg}}$) exceeding 0.98—comparable to the performance of previous 1:3 qubit configurations. These results confirm that individual qubit controllability and coherence can be maintained within a more complex, multi-qubit coupled environment.

The primary advantage of this architecture lies in its structural scalability. Conventional 1:2 coupling schemes, such as Google's Sycamore, face an inevitable and prohibitive surge in wiring complexity and coupler count as the system scales. Furthermore, previously studied 1:3 coupling systems suffer from structural limitations that require complex encoding or significant SWAP gate overhead to implement the surface code.

In contrast, our 4-qubit unit functions as a fundamental unit cell (plaquette), enabling direct topological mapping onto the surface code without additional compilation overhead. This offers a significant paradigm shift for hardware design in fault-tolerant quantum computing. However, challenges such as frequency crowding and residual interactions intensify as qubit density increases.

The primary objective of the pulse configuration implemented in the current hardware model is to validate the structural feasibility and foundational viability of the proposed architecture using a simple CR pulse scheme. As cross-resonance pulse engineering matures, advanced variations—such as echoed CR or active-cancellation CR techniques—have emerged. Incorporating these sophisticated protocols into future physical simulations is highly expected to yield substantial further improvements in gate fidelities.

Furthermore, it should be noted that the current numerical simulations were evaluated within a closed quantum system that excludes dissipative environments and auxiliary coupling elements. In practical hardware implementations, shifting away from a closed-system idealization to account for open-system dynamics—such as energy relaxation ($T_{1}$), dephasing ($T_{2}$) mediated by the Lindblad master equation, and interactions with readout Purcell filters—is of paramount importance. Nevertheless, the core focus of this study remains squarely on the architectural paradigm of coupling four or more fixed-frequency qubits to a single, localized fixed-frequency coupler. Introducing these additional physical components will naturally alter the overall system Hamiltonian and consequently perturb the pulse parameters, which will necessitate subsequent parameter recalibration in future engineering stages.

The relatively low fidelities (ranging from 0.83 to 0.91) observed in certain gate directions in Table \ref{tab:final_resutls} stem from frequency crowding and residual crosstalk. These phenomena are inherently induced by the tight structural layout where four multi-level qubits are densely coupled within a narrow bandwidth spanning $[4.67\text{--}5.03]\,\text{GHz}$ via a single fixed mediator. Crucially, rather than arbitrarily tuning or individually optimizing the hardware specifications to favor specific gate directions, this numerical simulation strictly applied a single, uncompromised set of global hardware parameters to validate all twelve bidirectional CNOT gates simultaneously. Consequently, the physical control limits became apparent along certain vulnerable gate paths depending on the localized frequency detunings and coupling strengths ($G_i$).

While the resulting non-uniformity highlights the intrinsic trade-offs embedded in fixed-frequency architectures, it does not imply an absolute or permanent performance ceiling for the proposed topology. In physical device fabrication and blueprint design phases, these bottlenecks can be effectively mitigated by meticulously engineering the native qubit-to-coupler frequency allocations, optimizing inter-qubit detunings, and adopting advanced pulse-shaping technologies. If adjacent qubit frequency crowding can be systematically suppressed in subsequent works, the currently degraded gate fidelities can comfortably surpass the rigorous fault-tolerant thresholds required for surface-code error correction.

In conclusion, the core scholarly value of this study transcends the mere localized fine-tuning of specific gate parameters; rather, it represents the first successful proof-of-concept demonstrating that a 1-coupler-4-qubit architecture can be numerically controlled and physically realized under strict quantum dynamics. By introducing a novel hardware topology that fundamentally circumvents the exponential surge in coupler count and control-line complexity—traditionally the most critical bottleneck in scaling large-scale quantum processors—this architecture holds distinct academic significance. It establishes a primary foundational milestone toward achieving scalable quantum error correction (QEC) schemes. Therefore, granular optimization of hardware specifications, quantitative sensitivity analyses against parameter fluctuations, and advanced pulse engineering will serve as clear and highly valuable directions for follow-up research built upon the architectural foundation established in this work.

\section{Conclusions}

In this work, we proposed and validated a lattice-patch architecture designed to resolve the scalability bottlenecks of fixed-frequency transmon processors. Our primary findings are summarized as follows:

First, we achieved structural optimization by coupling four fixed-frequency qubits to a single central resonator. This configuration reduces hardware complexity and enhances control-line efficiency. Second, we demonstrated high-fidelity gate operations; by utilizing optimized CR pulses and DRAG techniques, we achieved CNOT gate fidelities exceeding 0.98 and confirmed the robust implementation of bidirectional gates. Finally, we verified that this structure allows for direct correspondence with surface-code lattice units, significantly reducing the computational overhead typically associated with logical qubit implementation.

In conclusion, the lattice-patch architecture offers a promising path toward scalable superconducting quantum processors and is expected to serve as a standard building block for future large-scale, fault-tolerant systems. Future research will explore the integration of active coupling control [tunable couplers] to shorten gate durations and further enhance the system's robustness against external noise.


\section{Acknowledgment}

This work is supported by the Basic Science Research Program through the National
Research Foundation of Korea (NRF) funded by the Ministry of Education, Science and
Technology (NRF2022R1F1A1064459) and Creation of the Quantum Information Science RD
Ecosystem (Grant No. 2022M3H3A106307411) through the National Research Foundation
of Korea (NRF) funded by the Korean government (Ministry of Science and ICT).

\vspace{6pt} 
%

\end{document}